\documentclass[12pt]{article}
\usepackage{amssymb}
\usepackage[cp1251]{inputenc}
\usepackage[english]{babel}
\begin{document}

\begin{center}
{\Large {\bf Relativistic Dynamics of a Charged Particle in an Electroscalar Field}}

\vspace{1cm}

{\it  D.V. Podgainy$^1$, O.A. Zaimidoroga$^2$}

\vspace{0.5cm}

Joint Institute for Nuclear Research\\
141980, Dubna, Russia

$^1$ podgainy@jinr.ru\\

$^2$ zaimidor@jinr.ru \\

\end{center}

\vspace{1cm}

{\small 
This article devoted to relativistic dynamics of a charged massive particle in an electroscalar field. It represents a continuation of paper \cite{PoZa} where the authors constructed a non-relativistic theory which describes transverse electromagnetic waves along with longitudinal electroscalar ones, responsible for the wave transport of the Coulomb field. A new type of relativistic force exerted by electroscalar field on an electrically charged particle and the relativistic law of superposition of electromagnetic transverse and electroscalar longitudinal fields are established. Also, a relativistically invariant form of a Lagrangian describing the interaction between an electroscalar field and massive electrically charged particle is defined. 

}

\vspace{1cm}
{\it Keywords:} Maxwell electrodynamics, longitudinal electroscalar wave, transport of Coulomb field, relativistic dynamics
\vspace{0.5cm}

{PACS:} 03.50.De, 41.20.-q, 03.30.+p

\section*{Introduction}

The authors of paper \cite{PoZa} put forward a non-relativistic theory which describes the longitudinal electroscalar wave mode along with the transverse electromagnetic one, capable of propagating in vacuum. In the proposed theory, physical fields are of two types – vortex electric and magnetic fields as well as potential electric and 3-scalar ones. The fields of the first type can propagate in vacuum as a transverse electromagnetic wave while the fields of the second type as a longitudinal electroscalar wave where the vector of the electric field vibrates along the direction of propagation. It is the longitudinal electroscalar mode that performs transport of the Coulomb field, which is absent in the classical Maxwell theory. Absence of wave transport of the Coulomb field leads to certain obstacles in the canonical quantization of the Maxwell theory \cite{Dirac, FoPo, Polub} as well as to violation of the causality principle due to momentary propagation of the Coulomb field \cite{Chub}. The division of physical fields into vortex and potential ones is well known in the linear theory of elasticity \cite{LLTE}. Moreover, the linear elasticity theory is a fine mechanical analogy of the Maxwell theory, while perfect analogy is reached under the assumption that elastic continuum is incompressible \cite{Dubrovskii}. In other words, if potentials of the electromagnetic field are considered as elastic displacements in vacuum, the Maxwell theory describes the wave dynamics of absolutely incompressible continuum. From the standpoint of this mechanical analogy, the longitudinal electroscalar mode in the proposed by authors theory \cite{PoZa} is a consequence of “compressibility” of vacuum. It should be noted that a number of theoretical studies \cite{Waser, Vlaenderen, Khvorostenko} have been devoted to expansion of the classical Maxwell electrodynamics in the framework of the vector formalism, containing a longitudinal wave mode. However, such generalizations of the Maxwell theory have certain difficulties. One of the major ones is that in frames of the vector formalism presence of a longitudinal wave mode leads to the loss of gauge invariance of the theory and observability of electrodynamic potentials \cite{Rous}. The loss of gauge invariance, in its turn, entails problems related to electric charge conservation. The latter circumstance was demonstrated in the supplement to work \cite{PoZa} in terms of the Fock-Podolsky electrodynamics \cite{FoPo}, and, as was shown ibid, this theory can be presented in a gauge-invariant form but hereby the charges and currents become dependent on the gauge choice. The proposed by authors approach bases on the hypothesis that the Coulomb field, in the context of quantum-field concept, is a superposition of scalar photons, i.e. superposition of massless scalar particles with zero spin, rather than superposition of “time” photons \cite{Dirac, FoPo}. Constructively, this hypothesis is expressed through introduction of a 4-scalar potential along with the 4-vector one, with these two 4-potential fields being not connected with each other neither by differential nor any other correlations. The above circumstance allows preserve gauge invariance of transverse fields and avoid difficulties related to the law of electric charge conservation. 

In paper \cite{PoZa} there were established a non-relativistic force exerted by the electroscalar field on a charged particle and the non-relativistic law of superposition of electric vortex and electric potential fields. The present work is devoted to relativistic dynamics of a charged massive particle in an electroscalar field, namely, to the determination of the type of relativistic force and relativistic law of superposition of electromagnetic transverse and electroscalar longitudinal fields. Also, the relativistically invariant form of a Lagrangian describing the interaction of electroscalar field with a massive electrically charged particle is defined.

\section{Transformation properties of electroscalar fields}

The non-relativistic Lagrangian of interaction of a 4-scalar field $\lambda$ with a particle of charge $q$ and mass $m$, determined in paper \cite{PoZa}, has the form:
\begin{eqnarray}
L=\frac{m\textbf{v}^2}{2}-q\lambda,
\label{eq1}
\end{eqnarray}
while the density of the Lagrangian function for this field is defined by the expression
\begin{eqnarray}
{\cal L}_{EW}=\frac{W^2-\textbf{E}_{||}^2}{8\pi}+\rho\lambda=\frac{1}{8\pi}\left(\frac{1}{c}\frac{\partial\lambda}{\partial t}\right)^2-\frac{1}{8\pi}(\nabla\lambda)^2+\rho\lambda,
\label{eq2}
\end{eqnarray}
where $\rho$ is the density of electric charges; $c$ is the velocity of light in vacuum; and also the following definition for three-dimensional fields $\textbf{E}_{||}$ and $W$ are introduced:
\begin{eqnarray}
\textbf{E}_{||}=\nabla\lambda,\quad\quad W=-\frac{1}{c}\frac{\partial\lambda}{\partial t}.
\label{eq3}
\end{eqnarray}
The Lagrangian function (\ref{eq1}) leads to the following equation of particle motion:
\begin{eqnarray}
m\frac{d\textbf{v}}{dt}=-q\nabla\lambda=-q\textbf{E}_{||},
\label{eq4}
\end{eqnarray}
while for the 4-potential $\lambda$ we can obtain from (\ref{eq2}):
\begin{eqnarray}
-\frac{1}{c^2}\frac{\partial^2\lambda}{\partial t^2}+\Delta\lambda=-4\pi\rho.
\label{eq5}
\end{eqnarray}
Correspondingly, for 3-dimensional fields (\ref{eq3}) the following equations are derived:
\begin{eqnarray}\nonumber
&&\frac{1}{c}\frac{\partial\textbf{E}_{||}}{\partial t}+\nabla\,W=0,\\[0.2cm]\label{eq6}
&&\frac{1}{c}\frac{\partial W}{\partial t}+{\rm div}\textbf{E}_{||}=-4\pi\rho,\\[0.2cm]\nonumber
&&\textbf{rot E}_{||}=0.
\end{eqnarray}
The latter system of equations can be presented as a system of wave equations:
\begin{eqnarray}\label{eq7}
&&-\frac{1}{c^2}\frac{\partial^2\textbf{E}_{||}}{\partial t^2}+\Delta\textbf{E}_{||}=-4\pi\nabla\rho,\quad\quad \textbf{rot E}_{||}=0, \\[0.2cm]\nonumber
&&-\frac{1}{c^2}\frac{\partial^2 W}{\partial t^2}+\Delta W=\frac{4\pi}{c}\frac{\partial\rho}{\partial t}.
\end{eqnarray}
From the system of equations (\ref{eq6}) and (\ref{eq7}) it follows that longitudinal electroscalar waves are responsible for the transport of Coulomb field. In point of fact, the system of equations (\ref{eq6}) contains an equation for electrostatics ${\rm div}\textbf{E}_{||}=-4\pi\rho$, while the system (\ref{eq7}) has a solution in the form of longitudinal waves, the time-dependant heterogeneous density of electric charge being the source of these waves \cite{PoZa}. 

From the definition of three-dimensional fields $\textbf{E}_{||}$ and $W$ (\ref{eq3}) it follows that they are covariant components of the 4-vector:  
\begin{eqnarray}
{\cal W}_\mu=\partial_\mu\lambda=\left(\frac{1}{c}\frac{\partial\lambda}{\partial t}, \nabla_i\lambda\right)=(-W, \textbf{E}_{||})=({\cal W}_0, {\cal W}_i),
\label{eq8}
\end{eqnarray}
hereinafter the greek indices range over values from 0 to 3, while latin ones over values from 1 to 3; $\partial_\mu=(1/c \partial/\partial t, \nabla_i)$ are the components of a 4-gradient, and $\nabla_i$ are the components of a 3-gradient. In the Minkowski space with the metric tensor $\gamma_{\mu\nu}$ with the signature $(+1, -1, -1, -1)$, the square of the vector $W_\mu$ has the form:
\begin{eqnarray}
{\cal W}^2=\gamma^{\mu\nu}{\cal W}_\mu {\cal W}_\nu={\cal W}^\nu {\cal W}_\nu=W^2-\textbf{E}_{||}^2=\left(\frac{1}{c}\frac{\partial\lambda}{\partial t}\right)^2-\left(\nabla\lambda\right)^2,
\label{eq9}
\end{eqnarray}
hereinafter summation is supposed over repeated indices. Thus, Lagrangian of free electroscalar fields in four-dimensional notation can be represented as:
\begin{eqnarray}
{\cal L}_{\lambda}=\frac{1}{8\pi}{\cal W}_\nu {\cal W}^\nu=\frac{1}{8\pi}\left(\frac{1}{c}\frac{\partial\lambda}{\partial t}\right)^2-\frac{1}{8\pi}(\nabla\lambda)^2,
\label{lew}
\end{eqnarray}
while equations corresponding to this Lagrangian have the form:
\begin{eqnarray}\label{eq10}
&&\partial^\nu {\cal W}_\nu=0,\\[0.2cm]\nonumber
&&\partial_\mu {\cal W}_\nu-\partial_\nu {\cal W}_\mu=0.
\end{eqnarray}
Note that the latter equation in the above system, in terms of the definition (\ref{eq8}), is satisfied identically.

Under Lorenz transformations, components of the 4-vector ${\cal W}_\mu$ are transformed according to the law:
\begin{eqnarray}
{\cal W}_\mu'=\Lambda_\mu^\nu{\cal W}_\nu,
\label{eq11}
\end{eqnarray}
where $\Lambda_\mu^\nu$ is the matrix of Lorenz transformations $\Lambda_\mu^\nu$ \cite{Fok}:
\begin{eqnarray}
\Lambda_0^0=\frac{1}{\sqrt{1-\frac{V^2}{c^2}}},\quad\quad \Lambda_i^0=\frac{\Lambda_0^0}{c}V_i,\quad\quad \Lambda_i^k=\delta_i^k+(\Lambda_0^0-1)\frac{V_iV^k}{V^2},
\label{eq12}
\end{eqnarray}
while $V_i$ are the components of 3-velocity of relative motion of reference systems. Substituting (\ref{eq12}) into (\ref{eq11}), we obtain:
\begin{eqnarray}\label{eq13}
&&{\cal W}_0'=\frac{1}{\sqrt{1-\frac{V^2}{c^2}}}\left({\cal W}_0+\frac{1}{c}(V_k{\cal W}_k)\right)\\[0.2cm]\nonumber
&&{\cal W}_i'={\cal W}_i-\frac{V_i}{V^2}(V_k{\cal W}_k)+\frac{1}{\sqrt{1-\frac{V^2}{c^2}}}\frac{V_i}{V^2}\left((V_k{\cal W}_k)+\frac{V^2}{c}{\cal W}_0\right),
\end{eqnarray}
or in the three-dimensional notation
\begin{eqnarray}\label{eq14}
&&W'=\frac{1}{\sqrt{1-\frac{V^2}{c^2}}}\left(W-\frac{1}{c}(\textbf{V}\textbf{E}_{||})\right)\\[0.2cm]\nonumber
&&\textbf{E}_{||}'=\textbf{E}_{||}-\frac{\textbf{V}}{V^2}(\textbf{V}\textbf{E}_{||})+\frac{1}{\sqrt{1-\frac{V^2}{c^2}}}\frac{\textbf{V}}{V^2}\left((\textbf{V}\textbf{E}_{||})-\frac{V^2}{c}W\right).
\end{eqnarray}
From the first formula (\ref{eq14}) it follows that if in the unprimed reference system the field $\textbf{E}_{||}$ is absent, then
\begin{eqnarray}\nonumber
W'=\frac{W}{\sqrt{1-\frac{V^2}{c^2}}}.
\end{eqnarray}
Let us multiply the second formula (\ref{eq14}) scalarwise by $\textbf{V}$:
\begin{eqnarray}\nonumber
(\textbf{E}_{||}'\textbf{V})=\frac{1}{\sqrt{1-\frac{V^2}{c^2}}}\left((\textbf{V}\textbf{E}_{||})-\frac{V^2}{c}W\right).
\end{eqnarray}
In the case when $W=0$ in the initial reference system, the component of the vector $\textbf{E}_{||}$, directed along the velocity vector, changes according to the law:
\begin{eqnarray}\nonumber
\textbf{E}_{||}'=\frac{\textbf{E}_{||}}{\sqrt{1-\frac{V^2}{c^2}}}.
\end{eqnarray}
As this takes place, the component of the vector $\textbf{E}_{||}$, perpendicular to the velocity 
\begin{eqnarray}\nonumber
\textbf{E}_{||}-\frac{\textbf{V}}{V^2}(\textbf{V}\textbf{E}_{||}),
\end{eqnarray}
does not change under Lorenz transformations:
\begin{eqnarray}\nonumber
\textbf{E}_{||}'-\frac{\textbf{V}}{V^2}(\textbf{V}\textbf{E}_{||}')=\textbf{E}_{||}-\frac{\textbf{V}}{V^2}(\textbf{V}\textbf{E}_{||}).
\end{eqnarray}
Let us recall in brief transformation properties of transverse electromagnetic fields $\textbf{E}_\bot$ and $\textbf{H}$. As is known, 3-vector components of the Maxwell fields $\textbf{E}$ and $\textbf{H}$ behave under Lorenz transformations as components of antisymmetric second-rank tensor, namely, Minkowski tensor \cite{Fok}: 
\begin{eqnarray}
F_{\mu\nu} =\left( \begin{array}{cccr}
0&E_1&E_2&E_3\\
-E_1&0&-H_3&H_2\\
-E_2&H_3&0&-H_1\\
-E_3&-H_2&H_1&0\\
\end{array}\right)
\label{fmn}.
\end{eqnarray}
In this case, from the definition of the Maxwell electric field 
$$\textbf{E}=-\nabla\phi-\frac{1}{c}\frac{\partial \textbf{A}}{\partial t},$$
where $\phi$ and $\textbf{A}$ are the components of electromagnetic 4-vector potential, it follows that the field contains both the potential and vortex parts, however, in the Coulomb gauge $\phi=0, {\rm div}\textbf{A}=0$, the Maxwell electric field becomes a vortex one 
$${\textbf{E}_\bot}=-\frac{1}{c}\frac{\partial {\textbf{A}_\bot}}{\partial t}.$$
Note that the Maxwell field $\textbf{E}$ can be brought to the “vortex” type in any inertial reference system, despite the fact that Coulomb gauge conditions, $\phi=0, {\rm div}\textbf{A}=0$, are not Lorenz-invariant \cite{Dzhekson}. Thus, it is possible to write for the transformation of three-dimensional fields $\textbf{E}_\bot$ and $\textbf{H}$ the following:
\begin{eqnarray}\label{eq16}
\textbf{E}_\bot'=\frac{\textbf{V}}{V^2}(\textbf{V}\textbf{E}_\bot)+\frac{1}{\sqrt{1-\frac{V^2}{c^2}}}\left(\textbf{E}_\bot-\frac{\textbf{V}}{V^2}(\textbf{V}\textbf{E}_\bot)+\frac{1}{c}[\textbf{V}\times\textbf{H}]\right),\\[0.2cm]\nonumber
\textbf{H}'=\frac{\textbf{V}}{V^2}(\textbf{V}\textbf{H})+\frac{1}{\sqrt{1-\frac{V^2}{c^2}}}\left(\textbf{H}-\frac{\textbf{V}}{V^2}(\textbf{V}\textbf{H})-\frac{1}{c}[\textbf{V}\times\textbf{E}_\bot]\right).
\end{eqnarray}
From these formula it follows that $(\textbf{V}\textbf{E}_\bot')=(\textbf{V}\textbf{E}_\bot)$, i.e. the transverse electric field does not change along the velocity vector, moreover, for $\textbf{E}_\bot$ in the direction perpendicular to the velocity $\textbf{V}$ we obtain:
\begin{eqnarray}\nonumber
\left[\textbf{V}\times\textbf{E}_\bot'\right]=\frac{c\left[\textbf{V}\times\textbf{E}_\bot\right]+\left[\textbf{V}\times\left[\textbf{V}\times\textbf{H}\right]\right]}{c\sqrt{1-\frac{V^2}{c^2}}}.
\end{eqnarray}
In the case when magnetic field is absent in the unprimed reference system, the component of the electric field $\textbf{E}_\bot$, perpendicular to the velocity, follows the law: 
\begin{eqnarray}\nonumber
\left[\textbf{V}\times\textbf{E}_\bot'\right]=\frac{\left[\textbf{V}\times\textbf{E}_\bot\right]}{\sqrt{1-\frac{V^2}{c^2}}}\to \textbf{E}_\bot'=\frac{\textbf{E}_\bot}{\sqrt{1-\frac{V^2}{c^2}}}.
\end{eqnarray}
The same reasoning can be also applied for the magnetic field $\textbf{H}$. Thus, components of the transverse electric field $\textbf{E}_\bot$ and components of the longitudinal electric field $\textbf{E}_{||}$ under Lorenz transformations behave in opposite manner, which is absolutely natural as these three-dimensional fields are part of various four-dimensional objects, i.e. the field $\textbf{E}_\bot$ represents a component of the antisymmetric second-rank tensor, while $\textbf{E}_{||}$ defines the spatial part of the 4-vector. At the same time, in paper \cite{PoZa} there was established a non-relativistic law of superposition of electric fields: 
\begin{eqnarray}
\textbf{E}=\textbf{E}_\bot-\textbf{E}_{||}
\label{eq16}.
\end{eqnarray}
Apparently, this law will look differently in the relativistic case. It will be established in the next section.

\section{Lorenz-Invariant Interaction of Electroscalar Field with Electric Charge}

In paper \cite{PoZa}, an expression was determined for the force exerted by electroscalar (more precisely, electric) field on a particle with (rest) mass $m_0$ and charge $q$:
\begin{eqnarray}
\frac{d^2x_i}{dt^2}=-\frac{q}{m_0}{E_{||}}_i.
\label{eq17}
\end{eqnarray}
Moreover, the main peculiarity of the non-relativistic theory was, firstly, that electrostatic interaction was a particular case of the action of electroscalar field (see (\ref{eq6})), and, secondly, the 3-scalar field $W$ did not enter the expression for the force (\ref{eq17}). 

Next we assume that there exists such a frame of reference where 3-acceleration is proportional to $\textbf{E}_{||}$, i.e. there exists such a reference system where the charge rests and only electrostatic field can be found in this system:
\begin{eqnarray}
\frac{d^2x_i'}{dt^2}=-\frac{q}{m_0}{E_{||}}_i'.
\label{eq18}
\end{eqnarray}
For further consideration we also require formulas for 4-acceleration:
\begin{eqnarray}
a^0=\frac{1}{\sqrt{1-\frac{V^2}{c^2}}}\frac{d}{dt}\left(\frac{c}{\sqrt{1-\frac{V^2}{c^2}}}\right),\quad
a^i=\frac{1}{\sqrt{1-\frac{V^2}{c^2}}}\frac{d}{dt}\left(\frac{1}{\sqrt{1-\frac{V^2}{c^2}}}\frac{dx^i}{dt}\right),
\label{eq19}
\end{eqnarray}
and 4-velocities \cite{Fok}:
\begin{eqnarray}
u^0=\frac{c}{\sqrt{1-\frac{V^2}{c^2}}},\quad
u^i=\frac{1}{\sqrt{1-\frac{V^2}{c^2}}}\frac{dx^i}{dt}.
\label{eq20}
\end{eqnarray}
Components of 4-acceleration in the above-chosen reference system take the form:
\begin{eqnarray}
{a^0}',\quad {a^i}'=-\frac{q}{m_0}{{E_{||}}_i}',
\label{eq21}
\end{eqnarray}
in which case the instantaneous value of the 3-velocity in this reference system equals zero: ${u^0}'=c,\quad {u^i}'=0$. We can apply to (\ref{eq21}) formulas of Lorenz transformation:
\begin{eqnarray}
a^\mu=\Lambda^\mu_\nu {a^\nu}'\to a^0=\Lambda^0_k {a^k}'=-\frac{q}{m_0}\Lambda^0_k{{E_{||}}_k}',\quad a^i=\Lambda^i_k {a^k}'=-\frac{q}{m_0}\Lambda^i_k{{E_{||}}_k}'.
\label{eq22}
\end{eqnarray}
By substituting into the latter formula values for $\Lambda^\mu_\nu$ from (\ref{eq12}) and a value for ${{E_{||}}_i}'$, we obtain from (\ref{eq14}):
\begin{eqnarray}
&&a^0=-\frac{q}{m_0}\frac{c}{c^2-V^2}\left((\textbf{V}\textbf{E}_{||})-\frac{V^2}{c}W\right),\\[0.2cm]\nonumber
&&a^i=-\frac{q}{m_0}\left(\textbf{E}_{||}+\frac{\textbf{V}}{c^2-V^2}((\textbf{V}\textbf{E}_{||})-cW)\right)_i.
\label{eq23}
\end{eqnarray}
Taking into account that the velocity of the corresponding reference system $\textbf{V}$ coincides in this case with the velocity of motion of the charged particle $\textbf{v}$, we can write out the equations of motion which are valid for any inertial frame of reference:
\begin{eqnarray}\label{eq24}
&&a^0=-\frac{q}{m_0}\frac{c}{c^2-v^2}\left((\textbf{v}\textbf{E}_{||})-\frac{v^2}{c}W\right),\\[0.2cm]\nonumber
&&a^i=-\frac{q}{m_0}\left(\textbf{E}_{||}+\frac{\textbf{v}}{c^2-v^2}((\textbf{v}\textbf{E}_{||})-cW)\right)_i.
\end{eqnarray}
We will rewrite now the latter formulas in four-dimensional notation. For this purpose we introduce the antisymmetric second-rank tensor:
\begin{eqnarray}
{\cal W}_{\mu\nu}=\frac{1}{c}\left({\cal W}_\mu u_\nu-{\cal W}_\nu u_\mu\right).
\label{eq25}
\end{eqnarray}
In the three-dimensional notation the components of this tensor acquire the following form:
\begin{eqnarray}
{\cal W}_{0 i}=-{\cal W}_{i 0}=-\frac{v_iW+c{E_{||}}_i}{\sqrt{c^2-v^2}},\quad
{\cal W}_{ik}=-{\cal W}_{ki}=\frac{{E_{||}}_iv_k-{E_{||}}_kv_i}{\sqrt{c^2-v^2}}.
\label{eq26}
\end{eqnarray}
In a non-relativistic approximation, when $v/c\to 0$, the components of this tensor are represented as:
\begin{eqnarray}\nonumber
{\cal W}_{0 i}\approx {E_{||}}_i,\quad {\cal W}_{ik}\approx 0.
\end{eqnarray}
Let us define now the 4-vector: 
\begin{eqnarray}\nonumber
{\cal W}_{\mu\nu}u^\nu=\frac{1}{c}\left({\cal W}_\mu u_\nu u^\nu-{\cal W}_\nu u_\mu u^\nu\right)=c{\cal W}_\mu-\frac{u_\mu}{c}\left({\cal W}_0u^0+{\cal W}_ku^k\right). 
\end{eqnarray}
Calculation of the zero and spatial components of this vector yields:
\begin{eqnarray}\nonumber
&&{\cal W}_{0\nu}u^\nu=c{\cal W}_0-\frac{u_0}{c}\left({\cal W}_0u^0+{\cal W}_ku^k\right)=-cW+\frac{1}{1-\frac{v^2}{c^2}}\left(cW-(\textbf{v}\textbf{E}_{||})\right)=\\[0.2cm]\nonumber
&&-\frac{c^2}{c^2-v^2}((\textbf{v}\textbf{E}_{||})-cW)=\frac{m_0c}{q}a_0, 
\end{eqnarray}
\begin{eqnarray}\nonumber
&&{\cal W}_{i\nu}u^\nu=c{\cal W}_i-\frac{u_i}{c}\left({\cal W}_0u^0+{\cal W}_ku^k\right)=c{E_{||}}_i+\frac{v_ic}{c^2-v^2}\left(cW-(\textbf{v}\textbf{E}_{||})\right)=\\[0.2cm]\nonumber
&&c\left({E_{||}}_i+\frac{v_i}{c^2-v^2}\left(cW-(\textbf{v}\textbf{E}_{||})\right)\right)=\frac{m_0c}{q}a_i,
\end{eqnarray}
where $a_\mu$ are the covariant components of 4-acceleration:
$$a_\mu=\gamma_{\mu\nu}a^\nu.$$
Thus, in the four-dimensional notation system (\ref{eq24}) takes the form:
\begin{eqnarray}
a_\mu-\frac{q}{m_0c}{\cal W}_{\mu\nu}u^\nu=0.
\label{eq26a}
\end{eqnarray}
It can be seen from the latter formula that the role of intensity for electroscalar fields is played by the antisymmetric tensor ${\cal W}_{\mu\nu}$.

Recall that Lorenz-invariant equations of motion of a charged particle in Maxwell fields have the form \cite{LLTF}:
\begin{eqnarray}\label{eq26b}
&&\frac{d}{dt}\left(\frac{m_0c^2}{\sqrt{1-\frac{v^2}{c^2}}}\right)=q(\textbf{v}\textbf{E}),\\[0.2cm]\nonumber
&&\frac{d}{dt}\left(\frac{m_0\textbf{v}}{\sqrt{1-\frac{v^2}{c^2}}}\right)=q\left(\textbf{E}+\frac{1}{c}[\textbf{v}\times\textbf{H}]\right).
\end{eqnarray}
In the Coulomb gauge, taking into account the definition of the 4-acceleration (\ref{eq19}) and 4-velocity (\ref{eq20}), as well as the definition of the tensor $F_{\mu\nu}$ (\ref{fmn}), the latter system of equations can be presented in the 4-form: 
\begin{eqnarray}
a_\mu+\frac{q}{m_0c}F_{\mu\nu}^\bot u^\nu=0.
\label{eq27}
\end{eqnarray}
Here $F_{\mu\nu}^\bot$ is the Minkowski tensor in the Coulomb gauge:
\begin{eqnarray}\nonumber
F_{\mu\nu}^\bot=\frac{\partial A_\nu^\bot}{\partial x^\mu}-\frac{\partial A_\mu^\bot}{\partial x^\nu},\quad \phi=0,\quad {\rm div}\textbf{A}^\bot=0.
\end{eqnarray}
Summing up (\ref{eq26a}) and (\ref{eq27}), we obtain the covariant law of motion of a charged particle in external transverse electromagnetic and longitudinal electroscalar fields:
\begin{eqnarray}
a_\mu+\frac{q}{m_0c}O_{\mu\nu}u^\nu=0.
\label{eq28}
\end{eqnarray}
The tensor 
\begin{eqnarray}
O_{\mu\nu}u^\nu=F_{\mu\nu}^\bot-{\cal W}_{\mu\nu}
\label{eq29}
\end{eqnarray}
can be considered as a relativistic law of superposition of electromagnetic and electroscalar fields or as a generalized intensity. Components of this tensor in three-dimensional notation have the form: 
\begin{eqnarray}\nonumber
&&O_{0 i}=-O_{i 0}={E_\bot}_i-\frac{v_iW+c{E_{||}}_i}{\sqrt{c^2-v^2}},\\[0.2cm]\label{eq30}
&&O_{23}=-O_{32}=H_1+\frac{{E_{||}}_2v_3-{E_{||}}_3v_2}{\sqrt{c^2-v^2}},\\[0.2cm]\nonumber
&&O_{31}=-O_{13}=H_2+\frac{{E_{||}}_3v_1-{E_{||}}_1v_3}{\sqrt{c^2-v^2}},\\[0.2cm]\nonumber
&&O_{12}=-O_{21}=H_3+\frac{{E_{||}}_1v_2-{E_{||}}_2v_1}{\sqrt{c^2-v^2}}.
\end{eqnarray}
In a non-relativistic approximation at $v/c\to 0$ these formulas yield:
\begin{eqnarray}\nonumber
&&O_{0 i}\approx {E_\bot}_i-{E_{||}}_i,\\[0.2cm]\nonumber
&&O_{23}\approx H_1,\quad O_{31}\approx H_2,\quad O_{12}\approx H_3.
\end{eqnarray}
Thus, in a non-relativistic limit they represent the law of superposition for electric fields (\ref{eq16}) as does also the obtained in paper \cite{PoZa} expression for the total force acting on a charged particle. 

We return now to formula (\ref{eq26a}) and, using the definition for the 4-acceleration (\ref{eq19}), bring it into another form:
\begin{eqnarray}\nonumber
\frac{1}{\sqrt{1-\frac{v^2}{c^2}}}\frac{d}{dt}\left(m_0u_\mu\right)-\frac{q}{m_0c}{\cal W}_{\mu\nu}u^\nu=0.
\end{eqnarray}
In the latter formula we go over once again to the three-dimensional notation:
\begin{eqnarray}\label{eq31}
&&\frac{d}{dt}\left(\frac{m_0c^2}{\sqrt{1-\frac{v^2}{c^2}}}\right)=-\frac{q}{\sqrt{1-\frac{v^2}{c^2}}}\left((\textbf{v}\textbf{E}_{||})-\frac{v^2}{c}W\right),\\[0.2cm]\nonumber
&&\frac{d}{dt}\left(\frac{m_0\textbf{v}}{\sqrt{1-\frac{v^2}{c^2}}}\right)=-q\sqrt{1-\frac{v^2}{c^2}}\left(\textbf{E}_{||}+\frac{\textbf{v}}{c^2-v^2}\left((\textbf{v}\textbf{E}_{||})-cW\right)\right).
\end{eqnarray}
By summing up (\ref{eq26b}) and (\ref{eq31}), we obtain generalized expressions for the equations of motion of a charged particle in electromagnetic and electroscalar fields:
\begin{eqnarray}\label{eq32}
&&\frac{d}{dt}\left(\frac{m_0c^2}{\sqrt{1-\frac{v^2}{c^2}}}\right)=q\left[\textbf{E}_\bot-\frac{c}{\sqrt{c^2-v^2}}\left(\textbf{E}_{||}-\frac{\textbf{v}}{c}W\right)\right]\cdot\textbf{v},\\[0.2cm]\nonumber
&&\frac{d}{dt}\left(\frac{m_0\textbf{v}}{\sqrt{1-\frac{v^2}{c^2}}}\right)=q\left[\textbf{E}_\bot-\sqrt{1-\frac{v^2}{c^2}}\textbf{E}_{||}+\frac{1}{c}\left([\textbf{v}\times\textbf{H}]-\frac{\textbf{v}}{\sqrt{c^2-v^2}}\left((\textbf{v}\textbf{E}_{||})-cW\right)\right)\right].
\end{eqnarray}
In this regard, the first equation of the obtained system represents the law of conservation of the total kinetic energy of a charged particle, being a consequence of the second equation. Now let us multiply the second equation (\ref{eq32}) scalarwise by $\textbf{v}$:
\begin{eqnarray}
\textbf{v}\frac{d}{dt}\left(\frac{m_0\textbf{v}}{\sqrt{1-\frac{v^2}{c^2}}}\right)=q\left[(\textbf{v}\textbf{E}_\bot)-\sqrt{1-\frac{v^2}{c^2}}(\textbf{v}\textbf{E}_{||})-\frac{v^2}{c\sqrt{c^2-v^2}}\left((\textbf{v}\textbf{E}_{||})-cW\right)\right].
\label{eq33}
\end{eqnarray}
The left-hand side of this expression can be transformed using identical equation \cite{Fok}:
\begin{eqnarray}
a^\nu u_\nu\equiv 0\to \frac{d}{dt}\left(\frac{c^2}{\sqrt{1-\frac{v^2}{c^2}}}\right)\equiv \textbf{v}\frac{d}{dt}\left(\frac{\textbf{v}}{\sqrt{1-\frac{v^2}{c^2}}}\right).
\label{eq34}
\end{eqnarray}
The right-hand side of (\ref{eq33}) rearranges to:
\begin{eqnarray}\nonumber
&&q\left[(\textbf{v}\textbf{E}_\bot)-\sqrt{1-\frac{v^2}{c^2}}(\textbf{v}\textbf{E}_{||})-\frac{v^2}{c\sqrt{c^2-v^2}}\left((\textbf{v}\textbf{E}_{||})-cW\right)\right]=\\[0.2cm]\nonumber
&&q\left[(\textbf{v}\textbf{E}_\bot)-\sqrt{1-\frac{v^2}{c^2}}\left(\frac{c^2}{c^2-v^2}\left((\textbf{v}\textbf{E}_{||})-\frac{v^2}{c}W\right)\right)\right]=\\[0.2cm]\nonumber
&&q\left[(\textbf{v}\textbf{E}_\bot)-\frac{1}{\sqrt{1-\frac{v^2}{c^2}}}\left((\textbf{v}\textbf{E}_{||})-\frac{v^2}{c}W\right)\right].
\end{eqnarray}
Substituting the derived expression and (\ref{eq34}) into (\ref{eq33}), we obtain the first equation (\ref{eq32}). Certainly, this conclusion is also correct for each of the systems (\ref{eq26b}) and (\ref{eq31}). In other words, independent are only those equations which define the spatial parts of acceleration. As a consequence, from the Lagrangian formalism these particular equations are obtained. In the next section we will determine a relativistically invariant Lagrangian of interaction between an electrically charged particle and electroscalar field.

\section{Lorenz-Invariant Lagrangian of Electroscalar Field.}

In this section we consider a Lagrangian function which allows obtain from the variational principle the dynamic equation (\ref{eq31}). As for the Lagrangian of transverse fields, it was introduced in paper \cite{PoZa} and represents, as a matter of fact, a canonical Lagrangian for Maxwell electrodynamics (see, for example \cite{LLTF}), only in a Coulomb gauge. Note that imposing of Coulomb gauge on electromagnetic potentials is equivalent to the extraction from Maxwell electromagnetic fields of their transverse component. 

As a Lagrangian describing relativistic interaction of a massive charged particle with electroscalar field, we use the following expression:
\begin{eqnarray}
L=-m_0c^2\sqrt{1-\frac{v^2}{c^2}}\exp\left(\frac{q\lambda}{m_0c^2}\right).
\label{eq35}
\end{eqnarray}
This function describes multiplicative interaction of the scalar field $\lambda$ with a charged particle, i.e. the Lagrangian of a charged particle 
\begin{eqnarray}
L_0=-m_0c^2\sqrt{1-\frac{v^2}{c^2}}
\label{eq36}
\end{eqnarray}
multiplies by the function containing the field $\lambda$, in this case by the exponent:\\ $\exp(q\lambda/m_0c^2)$. The expression under exponent has a simple physical meaning, namely, it corresponds to the ratio of the potential energy of a particle in the field $q\lambda$ to the rest energy $m_0c^2$. Lagrangian (\ref{eq35}) can be given the form of a Lagrangian of a free particle (\ref{eq36}), supposing that the mass of a charged particle in the scalar field changes according to the law: 
\begin{eqnarray}
m_\lambda=m_0\exp\left(\frac{q\lambda}{m_0c^2}\right).
\label{eq37}
\end{eqnarray}
Note that such an exponential dependence of the inert mass on the field can be found in Nordstrom scalar theory of gravity \cite{Nord} (see also \cite{Serdukov}). Then (\ref{eq35}) acquires the form:
\begin{eqnarray}\nonumber
L_\lambda=-m_\lambda c^2\sqrt{1-\frac{v^2}{c^2}}.
\end{eqnarray}
We expand (\ref{eq35}) in powers of $v^2/c^2$:
\begin{eqnarray}\nonumber
L=-m_0c^2\left(1-\frac{v^2}{2c^2}\right)\exp\left(\frac{q\lambda}{m_0c^2}\right)=\left(\frac{m_0v^2}{2}-m_0c^2\right)\exp\left(\frac{q\lambda}{m_0c^2}\right),
\end{eqnarray}
and supposing that $q\lambda \ll m_0c^2$, expand into a series the exponent:
\begin{eqnarray}\nonumber
L=\left(\frac{m_0v^2}{2}-m_0c^2\right)\left(1+\frac{q\lambda}{m_0c^2}\right)\approx\frac{m_0v^2}{2}-q\lambda.
\end{eqnarray}
Thus, we arrive at the non-relativistic Lagrangian (\ref{eq1}), noting here that this Lagrangian is derived not only from the requirement of smallness of velocity of a charged particle $v \ll c$ but also from the requirement of smallness of interaction energy, as compared to the rest energy $q\lambda \ll m_0c^2$.

We calculate now the generalized momentum which corresponds to Lagrangian function (\ref{eq35}): 
\begin{eqnarray}
\textbf{P}=\frac{\partial L}{\partial\textbf{v}}=\frac{m_0\textbf{v}}{\sqrt{1-\frac{v^2}{c^2}}}\exp\left(\frac{q\lambda}{m_0c^2}\right)=\frac{m_\lambda\textbf{v}}{\sqrt{1-\frac{v^2}{c^2}}}.
\label{eq38}
\end{eqnarray}
The total time derivative of this expression is:
\begin{eqnarray}\label{eq39}
\frac{d}{dt}\frac{\partial L}{\partial\textbf{v}}&=&\exp\left(\frac{q\lambda}{m_0c^2}\right)\frac{d}{dt}\left(\frac{m_0\textbf{v}}{\sqrt{1-\frac{v^2}{c^2}}}\right)+\frac{m_0\textbf{v}}{\sqrt{1-\frac{v^2}{c^2}}}\frac{d}{dt}\left(\exp\left(\frac{q\lambda}{m_0c^2}\right)\right)=\\[0.2cm]\nonumber
&&\exp\left(\frac{q\lambda}{m_0c^2}\right)\left[\frac{d}{dt}\left(\frac{m_0\textbf{v}}{\sqrt{1-\frac{v^2}{c^2}}}\right)+\frac{q\textbf{v}}{c^2\sqrt{1-\frac{v^2}{c^2}}}\frac{d\lambda}{dt}\right].
\end{eqnarray}
Taking into account the expressions for the total time derivative of $\lambda$:
\begin{eqnarray}\nonumber
\frac{d\lambda}{dt}=\frac{\partial\lambda}{\partial t}+(\textbf{v}\nabla)\lambda=(\textbf{v}\textbf{E}_{||})-cW,
\end{eqnarray}
one can bring expression (\ref{eq39}) into the form:
\begin{eqnarray}
\frac{d}{dt}\frac{\partial L}{\partial\textbf{v}}=\exp\left(\frac{q\lambda}{m_0c^2}\right)\left[\frac{d}{dt}\left(\frac{m_0\textbf{v}}{\sqrt{1-\frac{v^2}{c^2}}}\right)+\frac{q\textbf{v}}{c^2\sqrt{1-\frac{v^2}{c^2}}}((\textbf{v}\textbf{E}_{||})-cW)\right].
\label{eq40}
\end{eqnarray}
Calculation of the gradient of (\ref{eq35}) leads to the expression:
\begin{eqnarray}
\frac{dL}{d\textbf{r}}=-q\sqrt{1-\frac{v^2}{c^2}}\exp\left(\frac{q\lambda}{m_0c^2}\right)\frac{d\lambda}{d\textbf{r}}=-q\sqrt{1-\frac{v^2}{c^2}}\exp\left(\frac{q\lambda}{m_0c^2}\right)\textbf{E}_{||}.
\label{eq41}
\end{eqnarray}
Substituting (\ref{eq40}) and (\ref{eq41}) into the Lagrangian equation
\begin{eqnarray}\nonumber
\frac{d}{dt}\frac{\partial L}{\partial\textbf{v}}=\frac{dL}{d\textbf{r}},
\end{eqnarray}
we obtain:
\begin{eqnarray}\nonumber
\frac{d}{dt}\left(\frac{m_0\textbf{v}}{\sqrt{1-\frac{v^2}{c^2}}}\right)+\frac{q\textbf{v}}{c^2\sqrt{1-\frac{v^2}{c^2}}}((\textbf{v}\textbf{E}_{||})-cW)=-q\sqrt{1-\frac{v^2}{c^2}}\textbf{E}_{||}.
\end{eqnarray}
Ultimately:
\begin{eqnarray}\nonumber
\frac{d}{dt}\left(\frac{m_0\textbf{v}}{\sqrt{1-\frac{v^2}{c^2}}}\right)=-q\sqrt{1-\frac{v^2}{c^2}}\left(\textbf{E}_{||}+\frac{\textbf{v}}{c^2-v^2}((\textbf{v}\textbf{E}_{||})-cW)\right).
\end{eqnarray}
Thus, Lagrangian function (\ref{eq35}) allows derive dynamic equation (\ref{eq31}). As it follows from this formula, the force $-q\textbf{E}_{||}$ has a zeroth order by $c$, being obtained in \cite{PoZa}, while the relativistic correction to this force is of the order $(v/c)^2$. 

Finally, we define the Hamiltonian corresponding to Lagrangian (\ref{eq35}):
\begin{eqnarray}
H=\textbf{P}\textbf{v}-L&=&\left(\frac{m_0v^2}{\sqrt{1-\frac{v^2}{c^2}}}+m_0c^2\sqrt{1-\frac{v^2}{c^2}}\right)\exp\left(\frac{q\lambda}{m_0c^2}\right)=\\[0.2cm]\nonumber
&&\frac{m_0c^2}{\sqrt{1-\frac{v^2}{c^2}}}\exp\left(\frac{q\lambda}{m_0c^2}\right)=\frac{m_\lambda c^2}{\sqrt{1-\frac{v^2}{c^2}}}.
\label{eq42}
\end{eqnarray}
So, the total energy of a charged particle in electroscalar field has the same form as the energy of a free particle, only with a changed mass $m_\lambda$. 

Now, we turn back to Lagrangian function (\ref{eq35}) and expand the exponent into a series up to terms of order $1/c^2$:
\begin{eqnarray}\nonumber
L=-m_0c^2\sqrt{1-\frac{v^2}{c^2}}-q\lambda\sqrt{1-\frac{v^2}{c^2}}.
\end{eqnarray}
In this formula we can go over from the Lagrangian to its density:
\begin{eqnarray}
{\cal L}=-\left(\mu_0c^2\sqrt{1-\frac{v^2}{c^2}}+\rho_0\lambda\sqrt{1-\frac{v^2}{c^2}}\right).
\label{eq43}
\end{eqnarray}
Here $\mu_0$ and $\rho_0$ define the mass and charge densities of a particle in the rest frame of the latter (we remind that the expressions $\mu_0\sqrt{1-v^2/c^2}$ and $\rho_0\sqrt{1-v^2/c^2}$ are Lorenz-invariant per se \cite{Fok}). It is also necessary to supplement the Lagrangian density ${\cal L}$ with a free Lagrangian (\ref{lew}) (more precisely, to subtract (\ref{lew}) from Lagrangian (\ref{eq43}) \cite{PoZa}): 
\begin{eqnarray}
{\cal L}=-\mu_0c^2\sqrt{1-\frac{v^2}{c^2}}-\rho_0\lambda\sqrt{1-\frac{v^2}{c^2}}-\frac{1}{8\pi}\left(\left(\frac{1}{c}\frac{\partial\lambda}{\partial t}\right)^2-(\nabla\lambda)^2\right).
\label{eq44}
\end{eqnarray}
Dropping from the above equation the kinematic term, we obtain a relativistically invariant Lagrangian:
\begin{eqnarray}
{\cal L}=\frac{1}{8\pi}\left(\left(\frac{1}{c}\frac{\partial\lambda}{\partial t}\right)^2-(\nabla\lambda)^2\right)+\rho_0\lambda\sqrt{1-\frac{v^2}{c^2}},
\label{lewr}
\end{eqnarray}
differing from (\ref{eq2}) by the factor $(1-v^2/c^2)^{1/2}$ at the summand $\rho_0\lambda$. The following field equations correspond to this Lagrangian:
\begin{eqnarray}
-\frac{1}{c^2}\frac{\partial^2\lambda}{\partial t^2}+\Delta\lambda=-4\pi\rho_0\sqrt{1-\frac{v^2}{c^2}}.
\label{eq45}
\end{eqnarray}
In the context of the definition for the 4-vector ${\cal W}_\mu$ (\ref{eq8}):
\begin{eqnarray}
&&\partial^\nu {\cal W}_\nu=-4\pi\rho_0\sqrt{1-\frac{v^2}{c^2}}\\[0.2cm]\nonumber
&&\partial_\mu {\cal W}_\nu-\partial_\nu {\cal W}_\mu=0.
\label{eq46}
\end{eqnarray}
Ultimately, for the 3-fields (\ref{eq3}) we obtain:
\begin{eqnarray}\nonumber
&&\frac{1}{c}\frac{\partial\textbf{E}_{||}}{\partial t}+\nabla\,W=0,\\[0.2cm]
&&\frac{1}{c}\frac{\partial W}{\partial t}+{\rm div}\textbf{E}_{||}=-4\pi\rho_0\sqrt{1-\frac{v^2}{c^2}},\\[0.2cm]\nonumber
&&\textbf{rot E}_{||}=0.
\end{eqnarray}
Note that the obtained relativistic corrections to the field equations (\ref{eq5}) and (\ref{eq6}) are of order $(v/c)^2$, i.e. the same as the order of relativistic correction to the force (\ref{eq17}).

\section*{Conclusion}

In conclusion we summarize and briefly discuss the obtained here results. It is appropriate to begin with the derived expression (\ref{eq31}) for the force exerted by electroscalar field on a charged particle. This formula significantly differs from its non-relativistic analogue gained in \cite{PoZa} and defined by expression (\ref{eq17}) in that the latter did not contain a three-dimensional scalar field $W$. This is due to that the theory developed in \cite{PoZa} can be considered as a post-Newtonian approximation of order $c^{-1}$. In fact, the force exerted on a charged particle by vortex electromagnetic fields (“transverse” component of the Lorenz force) is of order $c^{-1}$, which is clearly seen if this force is introduced via the potential $\textbf{A}_\bot$: 
\begin{eqnarray}\nonumber
\textbf{F}_\bot=\frac{q}{c}\left(-\frac{\partial \textbf{A}_\bot}{\partial t}+[\textbf{v}\times\textbf{rot}\textbf{A}_\bot]\right).
\end{eqnarray}
At the same time, the relativistic correction to the “longitudinal” force
\begin{eqnarray}\nonumber
\textbf{F}_{||}=-q\nabla\lambda
\end{eqnarray}
is of order $c^{-2}$:
\begin{eqnarray}\nonumber
-q\frac{\textbf{v}}{c^2}\left(1-\frac{v^2}{c^2}\right)^{-1/2}((\textbf{v}\textbf{E}_{||})-cW)=-q\left(1-\frac{v^2}{c^2}\right)^{-1/2}\frac{\textbf{v}}{c^2}\frac{d\lambda}{dt},
\end{eqnarray}
directed along the velocity vector of charged-particle motion and proportional to the total time derivative of the field $\lambda$.

One more remarkable peculiarity of the obtained results is the relativistic law of summation of transverse electromagnetic and longitudinal electroscalar fields (\ref{eq29})-(\ref{eq30}). It can be also regarded as a “kinematic induction” effect – from the above formulas it follows, for example, that the magnetic field is induced not only by the transverse electric field $\textbf{E}_\bot$ but also by the combination $[\textbf{v}\times\textbf{E}_{||}]$. In this light the magnetic field can be considered as a compensating field which arises at the motion of charge.

Finally, it is important to mention the derived here electroscalar Lagrangian (\ref{eq35}), which can be represented by a Lagrangian of a free particle, only with the inert mass depending on the field (\ref{eq37}). In this context it is of interest to examine the scalar theory of gravity proposed in \cite{Serdukov}. The Lagrangian for scalar gravity in this work has the form:
\begin{eqnarray}\nonumber
L_{G}=-m_0c^2\sqrt{1-\frac{v^2}{c^2}}\exp\left(\frac{\Phi}{c^2}\right),
\end{eqnarray}
where $\Phi$ is understood as a 4-scalar gravitational potential. As was mentioned above, the exponent included into electroscalar Lagrangian (\ref{eq35}) has a simple physical meaning – ratio of the potential energy of interaction to the rest energy of the particle, therefore, the developed by authors approach and advanced in \cite{Serdukov} approach to scalar gravity unite naturally:  
\begin{eqnarray}
L_{EW-G}=-m_0c^2\sqrt{1-\frac{v^2}{c^2}}\exp\left(\frac{q\lambda+m_0\Phi}{m_0c^2}\right),
\label{eq47}
\end{eqnarray}
i.e. the expression under exponent is the ratio of the total energy of interaction to the rest energy of the particle. In this light, an electrically charged massive particle possesses two degrees of freedom, described by corresponding four-scalar fields.

In closing it should be noted that currently there is no rigorous experimental proof of presence of the longitudinal electroscalar mode in Nature. Nevertheless, a number of works, in the opinion of the authors, contain clear evidence for its presence \cite{Monstein, Ignatiev}. It is also necessary to mention paper \cite{Lars} which provides experimental data on the manifestation of “longitudinal” forces in various media. In this context, it should be noted that the derived in this work relativistic correction to the electroscalar force is directed along the velocity vector of a particle, i.e. in frames of the electrodynamics of continuous media it is possible to introduce into the consideration a new type of “longitudinal” forces – forces acting in the direction of the electric current. Lastly, we point out that the elaborated by the authors theoretical approach, in particular the obtained Lagrangian (\ref{eq47}), can prove to be perspective when considering contemporary problems of cosmology – the problems of dark matter and dark energy (see, for example, papers \cite{Jose} and \cite{Gorbunov}, where the role of 4-scalar fields is discussed and an explanation of the problems of dark matter and dark energy is given).

\subsection*{Acknowlegements} 
This work carried out with partial support from the RFBR (Grant 11-01-00278a)

\end{document}